\documentstyle[11pt]{article}
\author{Adonai Sant'Anna\\Department of Mathematics - Federal University at Paran\'a\\P.O. Box 19081, Curitiba, PR, 81531-990, Brazil}
\title{Hidden variables, quasi-sets, and elementary particles}
\date{ }
\begin{document}
\maketitle

\newtheorem{definicao}{Definition}[section]
\newtheorem{teorema}{Theorem}[section]
\newtheorem{proposicao}{Proposition}[section]

\begin{abstract}
We have recently showed that it is possible to deal with collections of indistinguishable elementary particles (in the context of quantum mechanics) in a set-theoretical framework by using hidden variables, in a sense. In the present paper we use such a formalism as a model for quasi-set theory. Quasi-set theory, based on Zermelo-Fraenkel set theory, was developed for dealing with collections of indistinguishable but, in a sense, not identical objects.
\end{abstract}

\section{Introduction}

	We begin by considering that it is necessary to settle some philosophical terms in order to avoid confusions. When we say that $a$ and $b$ are {\em identicals\/}, we mean that they are the very {\em same\/} individual, that is, there are no `two' individuals at all, but only one which can be named indifferently by either $a$ or $b$. By {\em indistinguishability\/} we simply mean agreement with respect to attributes. We recognize that this is not a rigorous definition. Nevertheless such an intuition is better clarified in the next section.

	In physics, elementary particles that share the same set of state-in\-de\-pend\-ent (intrinsic) properties are usually refered to as {\it indistinguishable\/}. Although `classical particles' can share all their intrinsic properties, there is a sense in saying that they `have' some kind of {\em quid\/} which makes them individuals, since we are able to follow the trajectories of classical particles, at least in principle. That allows us to identify them. In quantum physics that is not possible, i.e., it is not possible, {\em a priori\/}, to keep track of individual particles in order to distinguish among them when they share the same intrinsic properties. In other words, it is not possible to label quantum particles.

	The problems regarding individuality of quantum particles have been discussed in recent literature by several authors. Some few of them are \cite{daCosta-94} \cite{DallaChiara-93} \cite{Krause-95b} \cite{Krause-98} \cite{Redhead-91} \cite{Sant'Anna-97} \cite{vanFraassen-91}. Many intrincate puzzles on the logical and philosophical foundations of quantum theory have been raised from these questions. For instance, there is the possibility that the collections of such entities may be not considered as sets in the usual sense. Yu. Manin \cite{Manin-76} proposed the search for axioms which should allow to deal with collections of indistinguishable elementary particles. Other authors \cite{DallaChiara-93} \cite{Krause-92} \cite{Krause-95b} have also considered that standard set theories are not adequate to cope with some questions regarding microphysical phenomena. These authors have emphasized that the ontology of microphysics apparently does not reduce to that one of usual sets, due to the fact that {\em sets\/} are collections of distinct objects.

	Quasi-set theory, based on Zermelo-Fraenkel set theory, was developed for dealing with collections of indistinguishable but, in a sense, not identical objects \cite{Krause-92}. Hence, quasi-set theory provides a mathematical background for dealing with collections of indistinguishable elementary particles as it has been shown in \cite{Krause-98}. In that paper, it has been shown how to obtain the quantum statistics into the scope of this non-standard approach.

	Nevertheless, it has been recently proposed that standard set theory is strong enough to deal with collections of {\em physically\/} indistinguishable quantum particles \cite{Sant'Anna-97} \cite{Sant'Anna-9*}, if we use some sort of hidden variable formalism. In the present paper we establish a connection between our hidden variable approach and quasi-set theory.

	Section 2 presents our hidden variable picture. Section 3 presents a very brief introduction for quasi-set theory and section 4 shows how to use the hidden variable formalism as a standard model for quasi-set theory. Finally, at section 5, we discuss some related lines of work.

\section{The Hidden Variable Formalism}

	Here we intend to show that it is possible to distinguish, at least in principle, among particles that are `{\em physically\/} indistinguishable', where by `physically indistinguishable' particles we mean, roughly speaking, those particles which share the same set of measurement values for their intrinsic properties.\footnote{By measurement values of intrinsic properties of a given particle we mean real numbers times an adequate unit of rest mass, charge, spin, etc., associated to the respective rest mass, charge, spin, etc. of this particle.} In a previous work \cite{Sant'Anna-97} we assumed that `{\em physically\/} indistinguishable particles' are those particles which have the same set of measurement values for a correspondent complete set of observables. It seems clear that such a modification simplifies our conceptual framework and it is still related to the usual understanding about the meaning of (physical) indistinguishability. A kind of distinction is possible if we consider each particle as an ordered pair whose first element is the mentioned set of measurement values of the intrinsic properties and the second element is a hidden property (a hidden variable) which intuitively corresponds to something which was not yet measured in laboratory. The mentioned `hidden property' does assume different values for each individual particle in a manner that it allows us to distinguish those particles which are in principle `physically' indistinguishable. Obviously that such a hidden property seems to have a metaphysical nature. Our proposed hidden variable hypothesis does have a metaphysical status. This is the kind of metaphysics that we advocate. The `reasonable' metaphysics should be that one which could provide a hope for a future new physics. This future new physics may correspond to more extended physical systems that are not, untill now, measured in laboratories.

	As remarked above, our concern here is only with the process of labeling physically indistinguishable particles. So, although we are not interested in describing here an axiomatic framework for quantum physics, quantum mechanics or even mechanics, we expect that our approach can be extended in order to encompass them. All that follows is performed in a standard set theory like Zermelo-Fraenkel with {\em Urelemente\/} (ZFU).\footnote{We interpret {\em Urelemente} as particles (in the sense of mechanics).}

	Our picture for describing indistinguishability issues in quantum physics is a set-theoretical predicate, following P. Suppes' ideas about axiomatization of physical theories \cite{Suppes-67}.

	Hence, our system has five primitive notions: $\lambda$, $X$, $P$, $m$, and $M$. $\lambda$ is a function $\lambda:N\rightarrow\Re$, where $N$ is the set $\{1,2,3...,n\}$, $n$ is a positive integer, and $\Re$ is the set of real numbers; $X$ and $P$ are finite sets; $m$ and $M$ are unary predicates defined on elements of $P$. Intuitivelly, the images $\lambda_{i}$ of the function $\lambda$, where $i\in N$, correspond to our hidden variable. We denote by $\Lambda_{N}$ the set of all $\lambda_{i}$, where $i\in N$. $X$ is a set whose elements should be intuitivelly interpreted as measurement values of the state-independent properties like rest mass, electric charge, spin, etc.. The elements of $X$ are denoted by $x$, $y$, etc. $P$ is to be interpreted as a set of particles. $m(p)$, where $p\in P$, means that $p$ is a microscopic particle, or a micro-object. $M(p)$ means that $p\in P$ is a macroscopic particle, or a macro-object. Actually, the distinction between microscopic and macroscopic objects, as mentioned here does not reflect, at least in principle, the great problem of explaining the distinguishability among macroscopic objects, since these are composed of physically indistinguishable things. As it is well known, Schr\"odinger explained that in terms of a {\em Gestalt\/} \cite{Schroedinger-52}. Nevertheless, this still remains as an open problem from the foundational (axiomatic) point of view. The following is a set-theoretical predicate for a system of ontologically distinguishable particles. We use the symbol `$=$' for the standard equality.\\

\noindent
{\bf Definition 2.1} {\it 
${\cal D_{O}} = \langle\lambda,X,P,m,M\rangle$ is a system of {\em ontologically distinguishable particles\/}, abbreviated as ${\cal D_O}$-system, if and only if the following six axioms are satisfied:
\begin{description}
\item [D1] $\lambda:N\rightarrow\Re$ is an injective function, whose set of images is denoted by $\Lambda_N$.
\item [D2] $P\subset X\times\Lambda_{N}$.
\end{description}

	{\rm We denote the elements of $P$ by $p,q,r,...$ when there is no risk of confusion.\\}

\noindent
{\bf Definition 2.2}
$\langle x,\lambda_{i}\rangle\doteq \langle y,\lambda_{j}\rangle\; \mbox{if, and only if,}\; x=y$.\\

\noindent
{\bf Definition 2.3}
If $p\in P$ and $q\in P$, we say that $p$ is {\em ontologically indistinguishable\/} from $q$ if, and only if, $p=q$, where $=$ is the usual equality between ordered pairs.\\

	{\rm The usual equality among ordered pairs $p = \langle x,\lambda_i\rangle\in P$ is a binary relation which corresponds to our ontological indistinguishability between particles, while $\doteq$ is another binary relation which corresponds to the physical indistinguishability between particles.

	Two particles are ontologically indistinguishable if and only if they share the same set of measurement values for their intrinsic physical properties and the same value for their hidden variables. Definition 2.3 says that two particles are physically indistinguishable if, and only if, they share the same set of measurement values for their intrinsic (physical) properties.}
\begin{description}
\item [D3] $(\forall x,y\in X)(\forall\lambda_i\in\Lambda_N)((\langle x,\lambda_{i}\rangle\in P \wedge \langle y,\lambda_{i}\rangle\in P) \rightarrow x = y).$
\item [D4] $(\forall p,q\in P)(M(p)\wedge M(q)\rightarrow (p\doteq q\rightarrow p=q))$.
\item [D5] $(\forall p,q\in P)(p\doteq q\wedge \neg(p=q)\to m(p)\wedge m(q)).$
\item [D6] $(\forall p\in P)((m(p)\vee M(p))\wedge\neg(m(p)\wedge M(p))$.
\end{description}
}

	Axiom {\bf D1\/} allows us to deduce that the cardinality of $\Lambda_{N}$ coincides with the cardinality of $N$ ($\#\Lambda_{N} = \# N$). Axiom {\bf D2\/} just says that particles are represented by ordered pairs\footnote{In \cite{daCosta-94} the authors discuss the possible representation of quantum particles by means of ordered pairs $\left< E,L\right>$, where $E$ corresponds to a predicate which in some way characterizes the particle in terms, e.g., of its rest mass, its charge, and so on, while $L$ denotes an apropriate label, which could be, for example, the location of the particle in space-time. Then, even in the case that the particles (in a system) have the same $E$, they might be distinguished by their labels. But if the particles have the same label, the tools of classical mathematics cannot be applied, since the pairs should be identified. In order to provide a mathematical distinction between particles with the same $E$ and $L$, these authors use quasi-set theory \cite{Krause-92} \cite{Krause-95b}. In the present picture, according to axioms {\bf D1}-{\bf D3}, it is prohibited the case where two particles have the same (ontological) label.}, where the first element intuitivelly corresponds to measurement values of all the intrinsic physical properties, while the second element corresponds to the hidden inner property that allows us to distinguish particles at an ontological level. Yet, axioms {\bf D2\/} and {\bf D3\/} guarantee that two particles that share the same values for their hidden variable are the very same particle, since our structure is set-theoretical and the equality $=$ is the classical one. Axiom {\bf D4\/} says that macroscopic objects that are physically indistinguishable, are necessarily identicals. Axiom {\bf D5\/} says that two particles physically indistinguishable that are not ontologically indistinguishable (they are ontologically distinguishable) are both microscopic particles. Axiom {\bf D6\/} means that a particle is either microscopic or macroscopic, but not both.

	Axiom {\bf D4\/} deserves further explanation. Let us observe that it was not postulated the existence of (in particular) micro-objects; but the axiomatic is compatible with such an hypothesis. Axiom {\bf D4\/} entails that (ontologically) distinct macro-objects are always distinguished by a measurement value; if two particles are macro-objects, then there exists a value for a measurement which distinguish them. Then, macro-objects, in particular, obey Leibniz's Principle of the Identity of Indiscernibles and we may say that (according to our axiomatics) classical logic holds with respect to them while micro-objects may be physically indistinguishable without the necessity of being `the same' object.

	In \cite{Sant'Anna-97} the axiomatic framework for a system of ontologically distinguishable particles is a little different from the present formulation. The main difference is on axiom {\bf D5\/}, which does not exist in \cite{Sant'Anna-97}. Such an axiom is necessary to prove the theorem that we present in the next subsection.

	We discuss in \cite{Sant'Anna-97} how our approach is out of the range of the well known proofs on the impossibility of hidden variables in the quantum theory, like von Neumann's theorem, Gleason's work, Kochen and Specker results, Bell's inequalities or other works where it is sustained that no distribution of hidden variables can account for the statistical predictions of the quantum theory \cite{Bohm-95}.

\section{Quasi-set theory}

The quasi-set theory ${\cal Q}$ is based on Zermelo-Fraenkel-like axioms and allows the presence of two sorts of atoms ({\it Urelemente\/}), termed $m$-atoms and $M$-atoms.\footnote{All the details of this section may be found in \cite{Krause-97}.} Concerning the $m$-atoms, a weaker `relation of indistinguishability' (denoted by the symbol $\equiv$), is used instead of identity, and it is postulated that $\equiv$ has the properties of an equivalence relation. The predicate of equality cannot be applied to the $m$-atoms, since no expression of the form $x = y$ is a formula if $x$ or $y$ denote $m$-atoms. Hence, there is a precise sense in saying that $m$-atoms can be indistinguishable without being identical. This justifies what we said above about the `lack of identity' to some objects.

	The universe of ${\cal Q}$ is composed by $m$-atoms, $M$-atoms and {\it quasi-sets\/} (qsets, for short). The axiomatics is adapted from that of ZFU (Zermelo-Fraenkel with {\it Urelemente\/}), and when we restrict the theory to the case which does not consider $m$-atoms, quasi-set theory is essentially equivalent to ZFU, and the corresponding quasi-sets can then be termed `ZFU-sets' (similarly, if also the $M$-atoms are ruled out, the theory collapses into ZFC, i.e., Zermelo-Fraenkel $+$ axiom of choice). The $M$-atoms play the role of the {\it Urelemente\/} in the sense of ZFU.

	The specific symbols of ${\cal Q}$ are three unary predicates $m$, $M$ and $Z$, two binary predicates $\equiv$ and $\in$ and an unary functional symbol $qc$. Terms and (well-formed) formulas are defined in the standard way, as are the concepts of free and bound variables, etc.. We use $x$, $y$, $z$, $u$, $v$, $w$ and $t$ to denote individual variables, which range over quasi-sets (henceforth, qsets) and {\em Urelemente}. Intuitively, $m(x)$ says that `$x$ is a microobject' ($m$-atom), $M(x)$ says that `$x$ is a macroobject' ($M$-atom) while $Z(x)$ says that `$x$ is a set'. The term $qc(x)$ stands for `the quasi-cardinal of (the qset) $x$'. The {\em sets} will be characterized as exact copies of the sets in ZFU. We also define that $x$ is a quasi-set, i.e., $Q(x)$ if, and only if, $x$ is neither an $m$-atom nor an $M$-atom.

	In order to preserve the concept of identity for the `well-behaved' objects, an {\it Extensional Equality\/} is introduced for those entities which are not $m$-atoms, on the following grounds: for all $x$ and $y$, if they are not $m$-atoms, then $$x =_{E} y\;\mbox{if, and only if,}\;$$
$$(Q(x)\wedge Q(y)\wedge\forall z ( z \in x \leftrightarrow z \in y )) \vee (M(x) \wedge M(y) \wedge x \equiv y)$$

	It is possible to prove that $=_{E}$ has all the properties of classical identity and so these properties hold regarding   $M$-atoms and `sets' (see below). In this paper, all references to `$=$' stand for `$=_E$', and similarly `$\leq$' and `$\geq$' stand, respectively, for `$\leq_E$' and `$\geq_E$'. Among the specific axioms of ${\cal Q}$, few of
them deserve explanation.  The other axioms are adapted from ZFU.

	For instance, to form certain elementary quasi-sets, such as those containing `two' objects, we cannot use something like the usual `pair axiom', since its standard formulation pressuposes identity; we use the weak relation of indistinguishability instead:

\vspace{0.3cm}
\noindent
[{\em The `Weak-Pair' Axiom\/}] For all $x$ and $y$, there exists a quasi-set whose elements are the indistinguishable objects from either $x$ or $y$. In symbols,\footnote{In all that follows, $\exists_Q$ and $\forall_Q$ are the quantifiers relativized to quasi-sets.} $$\forall x \forall y \exists_{Q} z \forall t (t \in z \leftrightarrow t \equiv x \vee t \equiv y)$$

	Such a quasi-set is denoted by $[x, y]$ and, when $x \equiv y$, we have $[x]$ by definition. We remark that this quasi-set {\it cannot\/} be regarded as the `singleton' of $x$, since its elements are {\it all\/} the objects indistinguishable from $x$, so its `cardinality' (see below) may be greater than $1$. A concept of {\it strong singleton\/}, which plays an important role in the applications of quasi-set theory, may be defined, as we shall mention below.

In ${\cal Q}$ we also assume a Separation Schema, which intuitivelly says that from a quasi-set $x$ and a formula $\alpha(t)$, we obtain a sub-quasi-set of $x$ denoted by $$[t\in x : \alpha(t)].$$

	We use the standard notation with `$\{$' and `$\}$' instead of `$[$' and `$]$' only in the case where the quasi-set is a {\it set\/}.

	It is intuitive that the concept of {\it function\/} cannot also be defined in the standard way, so we introduce a weaker concept of {\it quasi-function\/}, which maps collections of indistinguishable objects into collections of indistinguishable objects; when there are no $m$-atoms involved, the concept is reduced to that of function as usually understood. Relations, however, can be defined in the usual way, although no order relation can be defined on a quasi-set of indistinguishable $m$-atoms, since partial and total orders require antisymmetry, which cannot be stated without identity. Asymmetry also cannot be supposed, for if $x \equiv y$, then for every relation $R$ such that $\langle x, y \rangle \in R$, it follows that $\langle x, y \rangle = [[x]] = \langle y, x \rangle \in R$, by force of the axioms of ${\cal Q}$.\footnote{We remark that
$[[x]]$ is the same ($=_{E}$) as $\langle x, x \rangle$ by the Kuratowski's definition.}

	It is possible to define a translation from the language of ZFU into the language of ${\cal Q}$ in such a way that we can obtain a `copy' of ZFU in ${\cal Q}$. In this copy, all the usual mathematical concepts (like those of cardinal, ordinal, etc.) can be defined; the `sets' (in reality, the qsets which are `copies' of the ZFU-sets) turn out to be those quasi-sets whose transitive closure (this concept is like the usual one) does not contain $m$-atoms.

	Although some authors like Weyl \cite{Weyl-49} sustain that (in what regard cardinals and ordinals) ``the concept of ordinal is the primary one'', quantum mechanics seems to present strong arguments for questioning this thesis, and the idea of presenting collections which have a cardinal but not an ordinal is one of the most basic pressupositions of quasi-set theory.

	The concept of {\it quasi-cardinal\/} is taken as primitive in ${\cal Q}$, subject to certain axioms that permit us to operate with quasi-cardinals in a similar way to that of cardinals in standard set theories. Among the axioms for quasi-cardinality, we mention those below, but first we recall that in ${\cal Q}$, $qc(x)$ stands for the `quasi-cardinal' of the quasi-set $x$, while $Z(x)$ says that $x$ is a {\it set\/} (in ${\cal Q}$). Furthermore, $Cd(x)$ and $card(x)$ mean `$x$ is a cardinal' and `the cardinal of $x$' respectively, defined as usual in the
`copy' of ZFU we can define in ${\cal Q}$.

\vspace{0.3cm}
\noindent
[{\it Quasi-cardinality\/}] Every qset has an unique quasi-cardinal which is a cardinal (as defined in the `ZFU-part' of the theory) and, if the quasi-set is in particular a set, then this quasi-cardinal is its cardinal {\em stricto sensu\/}:\footnote{Then, every quasi-cardinal is a cardinal and
the above expression `there is a unique' makes sense. Furthermore, from the fact that the empty set $\emptyset$ is a set, it follows that its quasi-cardinal is 0.} $$\forall_{Q} x \exists_{Q} ! y (Cd(y) \wedge y = qc(x) \wedge (Z(x) \to y = card(x)))$$

	${\cal Q}$ still encompasses an axiom which says that if the quasi-cardinal of a quasi-set $x$ is $\alpha$, then for every quasi-cardinal $\beta \leq \alpha$, there is a subquasi-set of $x$ whose quasi-cardinal is $\beta$, where the concept of {\it subquasi-set\/} is like the usual one. In symbols,

\vspace{0.3cm}
\noindent
[{\it The quasi-cardinals of subquasi-sets\/}] $$\forall_{Q} x (qc(x) = \alpha \to \forall \beta (\beta \leq \alpha \to \exists_{Q} y (y \subseteq x \wedge qc(y) = \beta))$$

\vspace{3mm}
Another axiom states that

\vspace{0.3cm}
\noindent
[{\it The quasi-cardinal of the power quasi-set\/}]
$$\forall_{Q} x (qc({\cal P}(x)) = 2^{qc(x)})$$

\vspace{3mm}
\noindent
where $2^{qc(x)}$ has its usual meaning.

	As remarked above, in ${\cal Q}$ there may exist qsets whose elements are $m$-atoms only, called `pure' qsets. Furthermore, it may be the case that the $m$-atoms of a pure qset $x$ are indistinguishable from one another, in the sense of sharing the indistinguishability relation $\equiv$. In this case, the axiomatics provides the grounds for saying that nothing in the theory can distinguish among the elements of $x$. But, in this case, one could ask what it is that sustains the idea that there is more than one entity in $x$. The answer is obtained through the above mentioned axioms (among others, of course). Since the quasi-cardinal of the power qset of $x$ has quasi-cardinal $2^{qc(x)}$, then if $qc(x) = \alpha$, for every quasi-cardinal $\beta \leq \alpha$ there exists a subquasi-set $y \subseteq x$ such that $qc(y) = \beta$, according to the axiom about the quasi-cardinality of the subquasi-sets. Thus, if $qc(x) = \alpha \not= 0$, the axiomatics does not forbid the existence of $\alpha$  subquasi-sets of $x$ which can be regarded as `singletons'.

	Of course, the theory cannot prove that these `unitary' subquasi-sets (supposing now that $qc(x) \geq 2$) are distinct, since we have no way of `identifying' their elements, but qset theory is compatible with this idea.\footnote{The differences among such `unitary' qsets may perhaps be obtained from a distinction between `intensions' and `extensions' of concepts like `electron'. By this way we engage our approach into what Dalla-Chiara and Toraldo di Francia \cite{DallaChiara-93} termed the ``world of intensions''.} In other words, it is consistent with ${\cal Q}$ to maintain that $x$ has $\alpha$ elements, which may be regarded as absolutely indistinguishable objects. Since the elements of $x$ may share the relation
$\equiv$, they may be further understood as belonging to a same `equivalence class' (for instance, being indistinguishable electrons) but in such a way that we cannot assert either that they are identical or that they are distinct from one another (i.e., they act as `identical electrons' in the physicist's jargon).\footnote{The application of this formalism to the
concept of non-individual quantum particles has been proposed in \cite{Krause-95b}.}

	We define $x$ and $y$ as {\it similar\/} qsets (in symbols, $Sim(x,y)$) if the elements of one of them are indistinguishable from the elements of the another, that is, $Sim(x,y)$ if and only if $\forall z \forall t (z \in x \wedge t \in y \to z \equiv t)$. Furthermore, $x$ and $y$ are {\it Q-Similar\/} ($QSim(x,y)$) if and only if they are similar and have the same quasi-cardinality. Then, since the quotient qset $x/_{\equiv}$ may be regarded as a collection of equivalence classes of indistinguishable 
objects, then the `weak' axiom of extensionality is:

\vspace{0.3cm}
\noindent
[{\em Weak Extensionality\/}]
\begin{eqnarray}
\forall_{Q} x \forall_{Q} y (\forall z (z \in x/_{\equiv} \to \exists t 
(t \in y/_{\equiv} \wedge \, QSim(z,t)) \wedge \forall t(t \in
y/_{\equiv} \to\nonumber\\
\exists z (z \in  x/_{\equiv} \wedge \, QSim(t,z)))) \to x \equiv y)\nonumber
\end{eqnarray}

	In other words, the axiom says that those qsets that have `the same quantity of elements of the same sort\footnote{In the sense that they belong to the same equivalence class of indistinguishable objects.} are indistinguishable.

	Finally, let us remark that quasi-set theory is equiconsistent with standard set theories (like ZFC) (see \cite{Krause-95a}).

\section{A Set-Theoretical Model For Quasi-Sets in Terms of Hidden Variables}

	In this section we use $\cal D_{\cal O}$-system as a model for quasi-set theory. The interpretation is done according to the table given below:

\renewcommand{\arraystretch}{1.3}
\begin{equation}
\begin{array}{|c|c|}\hline
\mbox{Quasi-set theory} & \cal D_{\cal O}\mbox{-system} \\ \hline\hline
\mbox{Urelemente} & \mbox{elements of $P$}\\ \hline
\mbox{$p$ is a $m$-atom} & \mbox{$m(p)$}\\ \hline
\mbox{$p$ is a $M$-atom} & \mbox{$M(p)$}\\ \hline
\mbox{$Z(x)$} & \mbox{$x$ is set in ZFU}\\ \hline
\mbox{$qc(x)$} & \mbox{$card(x)$}\\ \hline
\mbox{$p=_Eq$} & p=q\\ \hline
\mbox{$p\equiv q$} & p\doteq q\\ \hline
\end{array}
\end{equation}

\vspace{3mm}

	In ${\cal D}_{\cal O}$-system we denote by $p_{\doteq}$ any set such that, ($\forall p\forall q$) ($p\in p_{\doteq}\wedge q\in p_{\doteq} \to p\doteq q$). Hence, the weak-pair in ${\cal D}_{\cal O}$-system corresponds to the {\em set \/} $\{p_{\doteq},q_{\doteq}\}$. With the table given above there is no difficulty to write the weak-pair axiom in ${\cal D}_{\cal O}$-system and to prove its translation (given by the table above) as a theorem of ZFU. The same occurs for the separation schema and the quasi-cardinality axioms. To prove the weak extensionality axiom we should consider another relation in ${\cal D}_{\cal O}$-system, which may be defined as follows:

\begin{definicao}

$P\doteq Q\;\mbox{if, and only if,}\;(\forall p\forall q)((p\in P\wedge q\in Q)\to p\doteq q)\wedge (card(P) = card(Q))$,

\end{definicao}

\noindent
where $card(X)$ stands for the standard cardinality of the set $X$.

	The proofs of the translations of the other axioms of quasi-set theory in ${\cal D}_{\cal O}$-system are not dif\-fi\-cult since ${\cal D}_{\cal O}$-system is defined as a set-theoretical predicate, and these proofs demand just simple properties of set-theory (ZFU). We do not give the details because there is no space in this book for a longer paper. But that is not a difficult task for the reader.

	As a final remark on this section, we note that it is not really necessary to interpret extensional equality, since this is given as a definition in quasi-set theory.

\section{Other Models and Related Questions}

	It is also possible to interpret quasi-sets in the set of rational numbers. We interpret the Urelemente in quasi-set theory as either rational numbers or nonconvergent Cauchy sequences of rational numbers.\footnote{We are obviously considering that the set of rational numbers is endowed with a metric. In our case, the metric is $d(x,y) = |x-y|$, where $x$ and $y$ are rational numbers.} To say that $x$ is an M-atom corresponds to say, in our proposed interpretation, that $x$ is a rational number. On the other hand, to say that $x$ is an m-atom means that $x$ is a nonconvergent Cauchy sequence of rational numbers. We say that two Cauchy sequences $(x_n)$ and $(y_n)$ are equivalent $((x_n)\sim (y_n))$ if, and only if, $z_n = (x_n - y_n)$ is a convergent sequence such that $z_n\to 0$. Such a binary relation $\sim$ is an equivalence relation. We say that two nonconvergent Cauchy sequences are {\em indistinguishable\/} if, and only if, they belong to the same equivalence class with respect to $\sim$. Extensional equality corresponds to the usual equality between rational numbers and sets of rational numbers. Quasi-cardinality may be interpreted as the usual cardinality in set-theory. And $Z(x)$ may be interpreted as `$x$ is a set. The proof that such an interpretation is a model for quasi-set theory is another task that we let as an exercise for the reader.

	We may also find non-standard interpretations for quasi-sets like the infinitesimals in non-standard analysis. Nevertheless, the main point we want to make in this paper is that quasi-sets are not so `weird', if we compare them with Zermelo-Fraenkel-like sets. If collections of elementary particles, in quantum physics, may be described by means of quasi-set theory, that does not mean that quantum world is completelly different from classical (macroscopic) world, at least from the mathematical point of view.

	Hence, we suggest the possibility of a complete classical picture for microscopic phenomena. Some authors have tried a similar way. Bohmiam mechanics is a well known example of a semi-classical picture for quantum mechanics \cite{Bohm-95}. Suppes and collaborators have also developed a particular description for some microscopic phenomena usually described by quantum physics \cite{Suppes-94} \cite{Suppes-96a} \cite{Suppes-96b}.

	One of the main points that makes quantum physics quite different from classical physics is the presence of nonlocal phenomena in the quantum world. Nevertheless, it is usually considered that the interference produced by two light beams, which is a nonlocal phenomenon, is determined by both their mutual coherence and the indistinguishability of the quantum particle paths. Mandel \cite{Mandel-91}, e.g., has proposed a quantitative link between the wave and the particle descriptions by using an adequate decomposition of the density operator. So, perhaps an adequate treatment for the problem of indistinguishability between elementary particle trajectories, in terms of hidden variables, may allow a classical picture for interference. Obviously, there is another nonlocal phenomenon, namely, Einstein-Podolsky-Rosen (EPR) experiment, which entails some fascinating results like teleportation \cite{Watson-97}. If we take very seriously the non individuality of quantum particles, space-time coordinates cannot be used to label elementary particles. Nevertheless, at the moment, we have no answer to the question if there is a relation between indistinguishability and nonlocality in the sense of EPR correlations.

\section{Acknowledgments}

	We acknowledge with thanks the suggestions and criticisms made by D\'ecio Krause.

\end{document}